\documentclass[aps]{revtex4} 

\usepackage{latexsym} 
\usepackage{amssymb} 
\usepackage{epsfig}    

\newcommand{\dbm}[1]{\hbox{\ensuremath{{#1}}}} 

\newcommand\prn[1]{\left(#1\right)}

\newcommand{\beq}{\begin{equation}}
\newcommand{\eeq}{\end{equation}}
\newcommand{\bea}{\begin{eqnarray}}
\newcommand{\eea}{\end{eqnarray}}

\def\slashchar#1{\setbox0=\hbox{$#1$}           
   \dimen0=\wd0                                 
   \setbox1=\hbox{/} \dimen1=\wd1               
   \ifdim\dimen0>\dimen1                        
      \rlap{\hbox to \dimen0{\hfil/\hfil}}      
      #1                                        
   \else                                        
      \rlap{\hbox to \dimen1{\hfil$#1$\hfil}}   
      /                                         
   \fi}

\newcommand{\refeq}[1]{Eq.~(\ref{#1})}

\newcommand{\refalg}[1]{Algorithm~(\ref{#1})}
\newtheorem{alg}{Algorithm}

\begin{document}

\preprint{NSF-KITP-08-52; CPHT - RR065.0908}
\title{Renormalised nonequilibrium quantum field theory: scalar fields}
\author{Sz. Bors\'anyi}
\email{email: s.borsanyi@sussex.ac.uk}
\affiliation{
Department of Physics and Astronomy, University of Sussex,
Brighton, East Sussex BN1 9QH, United Kingdom}
\altaffiliation[Part of this research has been carried out at]{
Kavli Institute for Theoretical Physics, UCSB,
Santa Barbara, CA 93106, USA.
}
\author{U. Reinosa}
\email{email: reinosa@cpht.polytechnique.fr}
\affiliation{Centre de Physique Th{\'e}orique, Ecole Polytechnique, CNRS,
92198, Palaiseau, France
}
\date{\today}


\begin{abstract}
We discuss the renormalisation of the initial value problem
in quantum field theory using the two-particle irreducible (2PI)
effective action formalism.
The nonequilibrium dynamics is renormalised by counterterms
determined in equilibrium. We emphasize the importance of the
appropriate choice of initial conditions and go beyond the Gaussian initial density operator by defining
self-consistent initial conditions. We study the corresponding time evolution and present a numerical
example which supports the existence of a continuum limit for this type of initial conditions.
\end{abstract}
\maketitle

\section{Introduction}

Nonequilibrium field theory is receiving an increasing level of
attention from the side of cosmologists as well as
from the heavy ion community.
The reheating of the postinflationary universe
\cite{Reheating}
the dynamics of symmetry breaking \cite{SymmetryBreaking} and
the formation and decay of cosmological defect networks \cite{Defects} as well
as the phase transitions in the early universe \cite{PhaseTransitions}
with possible relic gravitational waves \cite{GravityWaveProd}
are just some of the examples that require the study of out-of-equilibrium
fields in a cosmological context.
Similarly, the rapid thermalisation of the hot quark-gluon plasma
\cite{QGPThermalisation,HydroThermalisation}
and its driving mechanisms, such as the Weibel instability \cite{PlasmaInst}
raise questions in the realm of nonequilibrium field theory.

One of the simplest and most popular strategies to describe an
out-of-equilibrium field theory is the classical approximation. It has been
extensively used for reheating models of the early Universe
\cite{ClassicalReheating} and also for predicting the corresponding production
of gravitational waves \cite{ClassicalGravityWaves}. Other cosmological
applications include electroweak baryogenesis
\cite{EwBaryogenesis,EwBaryogensis2} as well as theories accomodating
non-fundamental strings \cite{ClassicalStrings} or domain walls
\cite{ClassicalWalls,BubblePreheating}, where the nonperturbative treatment is
essential.  The extreme excitation of the gluon field in a heavy ion collision
has also made the classical strategy applicable and very successful at early
times after a collision \cite{GlasmaLappi,GlasmaRomatschke}. 

The success of classical field theory indicates that many of the interesting
phenomena in high energy physics are actually classical. Indeed, genuine
quantum effects  play little role if the classical modes are
highly excited, and these classical fluctuations can play the role of
quantum particles. In order for the classical approach to work the UV modes
must remain unexcited to avoid Rayleigh-Jeans divergences. In fact, this
restricts the classical treatment to far-from-equilibrium settings, and
the classical dynamics automatically drives the system out of its 
range of validity in the course of equilibration.

It is still possible to split the momentum space of a theory into different
momentum regions, where hard degrees of freedom follow a quantum-mechanically
correct Hard Thermal Loop (HTL) dynamics to some finite perturbative order, while the infrared part follows the classical non-perturbative dynamics
\cite{HTLBodeker,HTLBratenPisarski}. This allows a kinetic description 
\cite{KineticHeinz,KineticMrowczynski} for
the hard modes in terms of a Vlasov equation \cite{CollectiveBlazotIancu}.
The interplay between classical waves and particles can give account for non-trivial dynamics, such as the development of plasma instabilities 
\cite{VlasovAM,VlasovBodeker}.
In this way one can avoid the problem of ultraviolet divergences, but the scale
separation is not always natural, especially if the coupling is not small.

A step towards the inclusion of quantum corrections from first principles is
the Hartree approximation \cite{HartreeLA,HartreedV}.  It assumes a constantly
Gaussian density operator and allows to account for quasi-particles propagating
in arbitrary inhomogeneous backgrounds, which can be as complicated as a
network of topological defects \cite{KinkHartree,DWHartree}.  Although it
completely neglects the scattering of the quasi-particles on each other,
non-perturbative particle creation mechanisms, like tachyonic instability
\cite{TachyonicPreheating} or parametric resonance \cite{ParametricPreheating}
are within its range of validity. Renormalisation in this framework has already
been discussed at length \cite{HartreeRenorm}.  Despite of its simplicity and
clean formulation the fact that thermalisation can not be described in this way
\cite{InhomHartree,InhomHartree2} explains why this approximation scheme could
not reach a wide acceptance.

The two-particle irreducible (2PI) effective action provides a first principles
approach to quantum field theory \cite{Baym,CJT}.
The systematic approximations, obtained from any  small
parameter expansion of the 2PI effective action, usually resum an entire series
of ladder diagrams which in turn play a particularly important role in solving
the secularity problem of out-of-equilibrium perturbation theory. This type of
resummation is also present in the Kadanoff-Baym equations \cite{KadanoffBaym}
as well as in Boltzmann equations \cite{CalzettaHu}.  It has been shown that
the first nontrivial truncation of the 2PI effective action beyond two-loop 
order already provides a sufficient framework to describe thermalisation in
scalar theories \cite{CoxBerges,AartsBergesScalarThermalisation,GreinerScalar,
ScalarIsotropisation,AmsterdamScalarThermalisation} as well as in a model with
fermions \cite{FermionThermalisation}.  The 2PI effective action has become a
standard framework for nonequilibrium quantum field theory
\cite{BergesIntroduction}, at present, with mostly scalar applications of
cosmological interest \cite{HeidelbergParametric,
AmsterdamTachyonic,DarmstadtFixedPoint}. 

The success of these practical applications has also encouraged more formal
investigations on the very foundations of the 2PI (and more generally $n$PI
\cite{nPI}) approach. To be considered as a sensible approach, the latter
should lead to approximations which reflect, as much as possible, the basic
properties of quantum field theories. The thermodynamical consistency as well
as the energy conserving nature of the time evolution have already been known
\cite{Formal2PI}. Understanding how global and local symmetries appear at the
level of the 2PI effective action has been studied in
\cite{HeesKnoll,Reinosa:2007vi}. Because some important aspects of  the
dynamics of hot gauge theories require a 3PI analysis \cite{LPM,ArnoldKinetic},
applications have also shifted focus to higher $n$PI effective actions
\cite{JurgennPI,Carrington4PI,AartsQED,Carrington3PI}. For low orders, however,
(e.g. for the truncation used in this paper) any higher $n$PI effective action
yields the equation of motion what one also finds in 2PI \cite{JurgennPI}.

Another important issue is renormalisation. So far, the latter has been considered in equilibrium for scalar
\cite{HeesKnoll,BlaizotIancuReinosa,HeidelbergRenorm,
CooperRenorm,JakovacRenorm,Szolt,JakovacRenormON}, fermionic
\cite{ReinsaFermions} as well as abelian gauge fields \cite{RSQEDRenorm}.
However, no similar studies exist so far out-of-equilibrium beyond
Hartree approximation, and most
applications were based on cut-off theories defined in terms of bare
parameters. In this work, we intend to fill this gap and show how
renormalisation results obtained in equilibrium could be used
out-of-equilibrium. For illustration, we consider a scalar
$(\lambda/4!)\varphi^4$ theory but our approach could be extended to other
theories of relevance.

In Sec.~\ref{sec:equilibrium}, we recall basic definitions and results in
equilibrium. Special attention is paid to the equivalence between different
contours in the truncated 2PI framework, which we use later on. In Sec.~\ref{sec:ooe} we move on to out-of-equilibrium situations and point out the relevance of beyond-Gaussian initial conditions for continuum field theories. We suggest a
self-consistent initial condition and study the corresponding time evolution in the 2PI three-loop approximation. Our numerical results strongly suggest the existence of a continuum limit for such initial conditions, unlike what happens with Gaussian initial conditions. Appendices \ref{app:alg} and \ref{app:isotropy}
 collect the algorithms used to obtain the results in Sec.~\ref{sec:ooe}.

\section{Equilibrium quantum field theory}\label{sec:equilibrium}
In this section we consider a real scalar field in equilibrium at a temperature
$\dbm{T=1/\beta}$. In this context, one is usually interested in determining
thermal expectation values of products of field operators ordered along a given
time contour $\mathcal{C}$. In Secs.~\ref{sec:contour_prop}-\ref{sec:equiv}, we
recall the definition of the contour-ordered propagator in equilibrium and
explain how to evaluate it non-perturbatively within the 2PI approximation
scheme. This requires non-perturbative renormalisation as we discuss in
Sec.~\ref{sec:renormalisation}.

\subsection{Contour-ordered propagator}\label{sec:contour_prop}
Consider a complex time contour $\mathcal{C}$, the properties of which we shall specify in what follows. The contour-ordered propagator is defined as
\begin{equation}\label{eq:contour_prop}
G(x,y)\equiv\Theta_{\mathcal{C}}(x_0,y_0)\,G^>(x,y)+\Theta_{\mathcal{C}}(y_0,x_0)\,G^>(y,x)\,,
\end{equation}
where $\Theta_{\mathcal{C}}(x_0,y_0)$ denotes the step function along the
contour, equal to one if $y_0$ precedes $x_0$ and zero in the opposite case.
The Wightman function\footnote{In the case of a complex scalar field one needs
to introduce two Wightman functions
${G^>(x,y)\equiv\langle\varphi(x)\varphi^\dagger(y)\rangle_\beta}$ and
$\dbm{G^<(x,y)\equiv\langle\varphi^\dagger(y)\varphi(x)\rangle_\beta}$. In
the present case, because the field is real, these two functions are related by
$\dbm{G^<(x,y)=G^>(y,x)}$.
}
 $G^>(x,y)$ appearing in Eq.~(\ref{eq:contour_prop}) is defined as
\begin{equation}\label{eq:Wigh}
G^>(x,y)\equiv\langle\varphi(x)\varphi(y)\rangle_\beta\equiv\frac{{\rm Tr}\,e^{-\beta H}\varphi(x)\varphi(y)}{{\rm Tr}\,e^{-\beta H}}\,.
\end{equation}
In equilibrium $G^>(x,y)$ depends on the difference of its arguments only and
we shall use this simplification when necessary. Using Lehmann's representation
one shows that $G^>(x-y)$ is analytic in the complex domain $\dbm{-\beta<{\rm
Im}\,(x_0-y_0)<0}$. From the equal-time commutation relations, we have
\begin{eqnarray}\label{eq:etcr}
\left[G^>(x,y)-G^>(y,x)\right]_{x_0=y_0} & \!\!\!=\!\!\! & 0\,,\\
\partial_{x_0}\left[G^>(x,y)-G^>(y,x)\right]_{x_0=y_0} & \!\!\!=\!\!\! & -i\delta^{(3)}(\vec{x}-\vec{y})\,.
\end{eqnarray}
Finally, one can easily show the KMS (Kubo-Martin-Schwinger) relation
\begin{equation}\label{eq:KMS_1}
G^>(x_0-i\beta,y_0;\vec{x},\vec{y})=G^>(y_0,x_0;\vec{y},\vec{x})\,,
\end{equation}
which holds for $\dbm{0<{\rm Im}\,(x_0-y_0)<\beta}$. In Fourier space,\footnote{We define the Fourier transform of $G^>(x)$ by $\dbm{G^>(p)\equiv\int d^4x \, e^{ipx} \, G^>(x)}$.} it reads
\begin{equation}\label{eq:KMS_2}
G^>(-p_0,-\vec{p})=e^{-\beta p_0}G^>(p_0,\vec{p})\,.
\end{equation}

\begin{figure}[htpb]
\centerline{\includegraphics[width=1.7in]{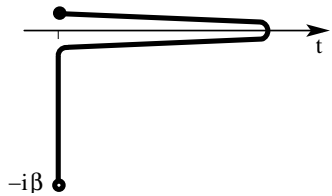}}
\caption{Close-time path. We here chose $\dbm{t_I=0}$.\label{fig:ctp}}
\end{figure}

From the analyticity property of $G^>(x,y)$ it follows that the contour-ordered
propagator $G(x,y)$ is properly defined only for contours with a decreasing
imaginary part. Moreover, when one tries to practically compute $G(x,y)$, a
second important constraint appears on the contour, namely that it should
stretch from some -- arbitrary -- initial time $t_I$ to a final time
$t_I-i\beta$. This condition, which appears both in the canonical and path
integral approaches, is dictated by the presence of the operator $e^{-\beta H}$
in the thermal average of Eq.~(\ref{eq:Wigh}). The same condition has been
recently found for the convergence  of real-time lattice simulations
\cite{Berges:2006xc}. 
One can construct numerous contours with decreasing imaginary part and stretching from time $t_I$ to $t_I-i\beta$. We shall refer to them as admissible contours. If one is interested in real-time aspects, the first branch of the contour should be contained in the real time axis. A particular example of such a contour is the close-time path, represented in Fig.~\ref{fig:ctp}. 

\subsection{Propagator from the 2PI approximation scheme}\label{sec:2PI}
Any propagator ordered along an admissible contour admits the path integral representation 
\begin{equation}\label{eq:path}
G(x,y)=\frac{\int\mathcal{D}\varphi\,\varphi(x)\varphi(y) \exp\left\{i\int_{\mathcal{C}} \mathcal{L}(x)\right\}}{\int\mathcal{D}\varphi\,\exp\left\{i\int_{\mathcal{C}} \mathcal{L}(x)\right\}}\,,
\end{equation}
where $\mathcal{L}(x)$ is the Lagrangian density
\begin{equation}
\mathcal{L}(x)=\frac{1}{2}(\partial\varphi(x))^2-\frac{1}{2}m^2(\varphi(x))^2-\frac{\lambda}{4!}(\varphi(x))^4
\end{equation}
 and the path integral is taken over fields satisfying the boundary condition $\dbm{\varphi(t_I;\vec{x})=\varphi(t_I-i\beta;\vec{x})}$. This path integral representation can be used to obtain a perturbative expansion of the contour-ordered propagator. It can also be used to define systematic non-perturbative approximations of the latter, as we now explain.

The contour-ordered propagator (\ref{eq:path}) can in fact be obtained from the generating functional
\begin{equation}
W[J,K]\equiv-i\ln\int\mathcal{D}\varphi\,\,e^{i\int_{\mathcal{C}}\mathcal{L}(x)+i\int_{\mathcal C}J(x)\varphi(x)+\frac{i}{2}\int_{\mathcal{C}}\int_{\mathcal{C}}\varphi(x)K(x,y)\varphi(y)}\,,
\end{equation}
either from a second derivative with respect to the source $J(x)$ or a single derivative with respect to the source $K(x,y)$, evaluated for vanishing sources. If we now introduce, the conjugate variables $\phi(x)$ and $G(x,y)$ defined by
\begin{equation}
\frac{\delta W[J,K]}{\delta J(x)}\equiv\phi(x)\quad{\rm and}\quad\frac{\delta W[J,K]}{\delta K(x,y)}\equiv\frac{1}{2}\Big[G(x,y)+\phi(x)\phi(y)\Big]\,,
\end{equation} 
we can construct the double Legendre transform of $W[J,K]$:
\begin{equation}
\Gamma[\phi,G] \equiv W[J,K] -\!\int_{\mathcal C}J(x)\phi(x)-\frac{1}{2}\int_{\mathcal C}\int_{\mathcal C} K(x,y)\,\Big[G(x,y)+\phi(x)\phi(y)\Big]\,,
\end{equation}
from which one can easily recover the initial sources $J(x)$ and $K(x,y)$ as
\begin{equation}
\frac{\delta\Gamma[\phi,G]}{\delta\phi(x)}=-J(x)\!-\!\!\int_{\mathcal{C}}K(x,y)\phi(y)\quad{\rm and}\quad\frac{\delta\Gamma[\phi,G]}{\delta G(x,y)}=-\frac{1}{2}K(x,y)\,.\label{eq:gapG}
\end{equation}
The benefit of this last equation is that, for a given source $K(x,y)$, it defines the two-point\footnote{Notice that $G(x,y)$ is the connected part of the contour-ordered propagator. In what follows we shall work in the situation where $\phi(x)$ vanishes and then both functions coincide.} function $G(x,y)$ as the solution to an implicit variational equation, which in turn is a powerful means for resumming Feynman diagrams and defining the contour-ordered propagator non-perturbatively. 

It is particularly convenient to introduce the decomposition
\begin{equation}
\Gamma[\phi,G]\equiv S_0[\phi]+\frac{i}{2}\mbox{Tr}_{\mathcal C}\,\ln G^{-1}+\frac{i}{2}\mbox{Tr}_{\mathcal{C}}\, G_0^{-1}\,G+\Gamma_{\rm int}[\phi,G]
\end{equation}
where $S_0[\varphi]$ denotes the free classical action along the contour $\mathcal{C}$ and $\dbm{G_0^{-1}(x,y)\equiv i(\partial^2_x+m^2)\delta_{\mathcal{C}}(x,y)}$ is the corresponding free inverse propagator. This decomposition allows one to rewrite Eq.~(\ref{eq:gapG}) as
\begin{equation}\label{eq:SD}
\delta_{\mathcal{C}}(x,y)=\int_{\mathcal{C}}d^4z\,\left[G_0^{-1}(x,z)-\Sigma(x,z)-iK(x,z)\right]G(z,y)\,,
\end{equation}
where the self-energy $\Sigma(x,y)$ is obtained from
\begin{equation}\label{eq:self}
\Sigma(x,y)=2i\frac{\delta\Gamma_{\rm int}[\phi,G]}{\delta G(y,x)}\,.
\end{equation}
Since $i\Gamma_{\rm int}[\phi,G]$ has a simple diagrammatic interpretation \cite{CJT} as the sum of all vac-to-vac two-particle-irreducible (2PI) diagrams of the shifted interaction part of the theory $S_{\rm int}[\phi+\varphi]$, it is very easy to define systematic non-perturbative approximations of the contour-ordered propagator using Eqs.~(\ref{eq:SD}) and (\ref{eq:self}). One has simply to select a certain number of 2PI diagrams in $\Gamma_{\rm int}[\phi,G]$, plug them in the right-hand-side of Eq.~(\ref{eq:self}) and solve the latter together with Eq.~(\ref{eq:SD}). In this work we concentrate on the three-loop approximation to $\Gamma_{\rm int}[\phi,G]$ in the absence of a field expectation value for which Eq.~(\ref{eq:self}) becomes 
\begin{equation}\label{eq:self_2loop}
\Sigma(x,y)\equiv-i\Sigma_0(x)\delta_{\mathcal{C}}(x,y)+\Theta_{\mathcal{C}}(x_0,y_0)\Sigma^>(x,y)+\Theta_{\mathcal{C}}(y_0,x_0)\Sigma^>(y,x)
\end{equation}
with
\begin{equation}
\Sigma_0(x)=\frac{\lambda}{2}\,G(x,x) \quad {\rm and} \quad \Sigma^>(x,y)=-\frac{\lambda^2}{6}\,G^>(x,y)^3\,.
\end{equation}

\subsection{Equivalence of contours}\label{sec:equiv}
Because we are discussing equilibrium, we set the source $K(x,y)$ to zero for the moment. We shall later consider, in Sec.~\ref{sec:ooe}, a non-vanishing source has a means to bring the system out-of-equilibrium.

\begin{figure}[htpb]
\centerline{\includegraphics[width=1.7in]{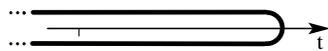}}
\caption{Real-time path.\label{fig:rtp}}
\end{figure}

Inverting equation Eq.~(\ref{eq:SD}) to obtain the propagator $G(x,y)$ in terms of the self-energy $\Sigma(x,y)$ is usually rendered cumbersome due to the non-real parts of the contour. As an important simplification, in App.~\ref{app:equivalence} we prove that, in equilibrium, any propagator ordered along an admissible contour can be reconstructed from the so-called real-time propagator, obtained from solving Eqs.~(\ref{eq:SD}) and (\ref{eq:self_2loop}) on the real-time path (depicted in Fig.~\ref{fig:rtp}), with the KMS and equal-time commutation relations as boundary conditions. 
The KMS condition is necessary because, in going from the close-time path to the real-time path, one looses any reference to the temperature and one needs a sensible way to reintroduce it.

In what follows, we restrict our analysis to the real-time path. In this case, Eq.~(\ref{eq:SD}) can be put in a more tractable form by introducing the functions
\begin{equation}
F(x,y)\equiv\frac{1}{2}\Big[G^>(x,y)+G^>(y,x)\Big]\,,\;\rho(x,y)\equiv i\Big[G^>(x,y)-G^>(y,x)\Big]
\end{equation}
as well as
\begin{equation}
\Sigma_F(x,y)\equiv\frac{1}{2}\Big[\Sigma^>(x,y)+\Sigma^>(y,x)\Big]\,,\;\Sigma_\rho(x,y)\equiv i\Big[\Sigma^>(x,y)-\Sigma^>(y,x)\Big]\,.
\end{equation}
 Using the equal-time commutation relations, one can then recast Eq.~(\ref{eq:SD}) into a pair of partial differential equations
\begin{eqnarray}
\prn{\partial_x^2+m^2+\Sigma_{0}}F(x) & \!\!\!\!=\!\!\!\! &
\!\!\int\limits^{0}_{-\infty} \!dz^4\Sigma_F(x-z)\rho(z)
-\!\!\int\limits^{x_0}_{-\infty} \!dz^4\Sigma_\rho(x-z)F(z)\,,\label{eq:Feq}\\
\prn{\partial_x^2+m^2+\Sigma_{0}}\rho(x) & \!\!\!=\!\!\! &
- \!\int\limits^{x_0}_0 \!dz^4\Sigma_\rho(x-z)\rho(z)\,,\label{eq:rhoeq}
\end{eqnarray}
where $\Sigma_0$, $\Sigma_F(x)$ and $\Sigma_\rho(x)$ can also be expressed in terms of $F(x,y)$ and $\rho(x,y)$. In writing these equations, we have used the fact that, in the absence of sources the functions $F(x,y)$, $\rho(x,y)$, $\Sigma_F(x,y)$ and $\Sigma_\rho(x,y)$ only depend on the difference of their arguments. 

In order to yield a particular solution, the previous two equations need to be
supplemented by some boundary conditions. Those for the spectral density
$\rho(x)$ are fixed by the equal time commutation relations:
$\dbm{\left.\rho(x)\right|_{x_0=0}=0}$ and $\dbm{\left.\partial_{x_0}\rho(x)\right|_{x_0=0}=1}$.  As for $F(x)$, the boundary
condition is nothing but the KMS condition we alluded to before. In terms of
$F/\rho$ components, it takes the form
\begin{equation}\label{eq:KMS}
F(p_0,\vec{p})=-i\left(\frac{1}{2}+f(p_0)\right)\rho(p_0,\vec{p})\,,
\end{equation}
where $\dbm{f(p_0)=1/(e^{\beta p_0}-1)}$ denotes the Bose-Einstein factor. In fact once the KMS condition is assumed, the system of equations (\ref{eq:Feq})-(\ref{eq:rhoeq}) becomes redundant. To see this in our particular example, notice that 
\begin{eqnarray}
\Sigma^>(p) & \!\!\!=\!\!\! & -\frac{\lambda^2}{6}\,\int\!\!\frac{d^4k}{(2\pi)^4}\!\int\!\!\frac{d^4l}{(2\pi)^4}\,G^>(k)\,G^>(l)\,G^>(p-k-l)\,,
\end{eqnarray}
from which one can check that $\Sigma^>(p)$, and in turn $\Sigma_F(p)$ and $\Sigma_\rho(p)$, obey KMS conditions similar to Eqs.~(\ref{eq:KMS_2}) and (\ref{eq:KMS}) respectively. Then writing Eqs.~(\ref{eq:Feq}) and (\ref{eq:rhoeq}) in Fourier space (we set $\dbm{\omega_{\vec{p}}^2\equiv \vec{p}^2+m^2+\Sigma_0}$):
\begin{eqnarray}
(-p_0^2+\omega_{\vec{p}}^2)\,F(p) & \!\!\! = \!\!\! &
\int \frac{d\omega}{2\pi}\left[
\frac{\Sigma_F(p)\,\rho(\omega;\vec p)}{i(p_0-\omega-i\epsilon)}
+
\frac{\Sigma_\rho(\omega;\vec p)\,F(p)}{i(p_0-\omega+i\epsilon)}
\right]\,,
\label{eq:fourierFeq}
\\
(-p_0^2+\omega_{\vec{p}}^2)\,\rho(p) & \!\!\!=\!\!\! &
\int \frac{d\omega}{2\pi}\left[
\frac{\Sigma_\rho(p)\,\rho(\omega;\vec p)}{i(p_0-\omega-i\epsilon)}
+
\frac{\Sigma_\rho(\omega;\vec p)\,\rho(p)}{i(p_0-\omega+i\epsilon)}
\right]\,,
\label{eq:fourierrhoeq}
\end{eqnarray}
it is straightforward to check that Eq.~(\ref{eq:fourierFeq}) implies Eq.~(\ref{eq:fourierrhoeq}) and vice versa. We conclude that in equilibrium, one only needs to solve Eq.~(\ref{eq:rhoeq}) for $\rho(x)$ and determine $F(x)$ from the KMS condition. Equation (\ref{eq:rhoeq}) needs to be renormalised, as we now explain.

\subsection{Renormalisation}\label{sec:renormalisation}

Equation (\ref{eq:rhoeq}) for the spectral density is usually plagued by ultraviolet divergences which one needs to renormalise in order to define a continuum limit. It has to be noted that, due to the presence of a Landau pole, this limit does not exist in the strict sense. It is true, however, that there is a wide range of couplings where the Landau pole is far in the UV and renormalisation can ensure the insensitivity to the cut-off, if it is significantly higher than other physical scales, but does not exceed the Landau pole, at which the computed bare coupling diverges. We use the term ``continuum limit'' in this restricted sense. Pattern of the divergence of bare parameters in the vicinity of the Landau pole has been discussed in \cite{RenormThermo} in the 2PI three-loop approximation, that we also use here. 

\begin{figure}[htpb]
\centerline{\includegraphics[width=0.7in]{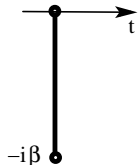}}
\caption{Imaginary time path.\label{fig:itp}}
\end{figure}

Renormalisation on the real-time path has been considered in \cite{HeesKnoll}. In this work we shall rather consider renormalisation in the so-called imaginary-time path \cite{BlaizotIancuReinosa} and infer renormalisation on the real-time path. Indeed, according to App.~\ref{app:equivalence}, the solution to Eqs.~(\ref{eq:SD}) and (\ref{eq:self_2loop}) on the real-time path supplemented by the KMS and equal-time commutation relations can be used to construct propagators ordered along any admissible contour, at the same level of approximation. In particular, if one chooses the imaginary-time path depicted in Fig.~\ref{fig:itp}, one obtains the so-called imaginary-time or Euclidean propagator ($\dbm{-\beta<\tau<\beta}$)
\begin{equation}
G_E(\tau;\vec{x})\equiv \Theta(\tau)G^>(-i\tau;\vec{x})+\Theta(-\tau)G^>(i\tau;-\vec{x})\,,
\end{equation}
where $G^>(-i\tau;\vec{x})$ is the analytic continuation of the real-time Wightman function to the imaginary-time path, as defined in Eq.~(\ref{eq:analytic}) of App.~\ref{app:equivalence}. 
The KMS condition implies that $G_E(\tau;\vec{x})$ is $\beta$-periodic in $\tau$ from which one concludes that $G_E(\tau;\vec{x})$ can equally be represented by its Fourier modes $G_E(p)$ with $\dbm{p\equiv(i\omega_n;\vec{p})}$ and $\dbm{\omega_n=(2\pi/\beta)n}$ a discrete Matsubara frequency:
\begin{equation}
G_E(i\omega_n;\vec{p})\equiv\int_0^\beta \!\!d\tau\!\int \!d^3x\,e^{i\omega_n\tau}e^{-i\vec{p}\cdot\vec{x}}G^>(-i\tau;\vec{x})\,.
\end{equation}
Using Eq.~(\ref{eq:analytic}) and the KMS condition, one then shows that
\begin{equation}\label{eq:rel}
G_E(i\omega_n;\vec{p})=\int\!\frac{d^4p}{(2\pi)^4}\frac{i\rho(p_0;\vec{p})}{i\omega_n-p_0}
\end{equation}
from which it follows that
\begin{equation}\label{eq:Im}
\rho(p_0;\vec{p})=2i\,{\rm Im}\,G_E(p_0+i\varepsilon;\vec{p})\,.
\end{equation}
We conclude from this, that in order to renormalise $\rho(p_0;\vec{p})$, it is
enough to renormalise $G_E(i\omega_n;\vec{p})$. This assumes that the analytic
continuation involved in Eq.~(\ref{eq:Im}) does not bring in new divergences. This assumption is plausible
since, in what follows, we restrict to a spatial cut-off and leave the time
direction in the continuum, see below.

\subsubsection{Imaginary time counterterms}
In Fourier space, the Euclidean propagator in the approximation at hand is such that $\dbm{G^{-1}_E(p)\equiv G_{E,0}^{-1}(p)+\Sigma_E(p)}$ with $\dbm{G_{E,0}^{-1}(p)\equiv \omega_n^2+\vec{p}^2+m^2}$ and
\begin{equation}\label{eq:SE}
\Sigma_E(p)=\delta\Sigma^{ct}_E(p)+\frac{\lambda}{2}\int_k\!\!\!\!\!\!\Sigma\,G_E(k)-\frac{\lambda^2}{6}\int_k\!\!\!\!\!\!\Sigma\int_l\!\!\!\!\!\Sigma\, G_E(k)G_E(l)G_E(p-k-l)\,,
\end{equation}
The piece $\delta\Sigma^{ct}_E(p)$ contains counterterms which should be
adjusted in such a way that UV divergences are absorbed.  Due to Eq.~(\ref{eq:rel}), the counterterms in imaginary-time are exactly those needed to renormalise the equations in real-time. Strictly speaking this is true if the time integral
is not discretized. For this reason we introduce a spatial cut-off only. For such an anisotropic regularisation and at three-loop order in
the 2PI-loop expansion the counterterm piece reads
\begin{equation}
\delta \Sigma^{ct}_E(p)=\delta Z_t\,\omega_n^2+\delta Z_s\,\vec{p}^2+ \delta M^2
\label{eq:dSigmaE2}
\end{equation}
with
\begin{eqnarray}
\delta M^2\equiv\delta m^2_E + \frac{\delta \lambda_E}{2} \,\int_k\!\!\!\!\!\!\Sigma\, G_E(k)\,.
\label{eq:dMsqr}
\end{eqnarray}
The three counterterms in Eq.~(\ref{eq:dSigmaE2}) can be fixed using renormalisation conditions at a reference temperature $T^\star$ and a reference momentum $p^\star$
\begin{equation}\label{eq:ren1}
\Sigma^\star_E(p^\star)=0\,,\qquad 
\left.\frac{d\Sigma^\star_E(p)}{d\omega_n^2}\right|_{p=p^\star}=0\,,
\qquad
\left.\frac{d\Sigma^\star_E(p)}{dp_3^2}\right|_{p=p^\star}=0\,,
\label{eq:RC1}
\end{equation}
see App.~\ref{app:alg} for Algorithm (\ref{alg:SigmaE}) which finds these counterterms. These three renormalisation conditions are, however, not sufficient to ensure insensitivity to the UV cut-off at a temperature $T$ away from the renormalisation temperature $T^\star$. For this to be achieved one needs to adjust the coupling counterterm $\delta\lambda_E$ in such a way that coupling UV subdivergences which appear in $\Sigma_E(p)$ are properly absorbed. These subdivergences can easily be accounted for by remarking that they are obtained after opening a perturbative line (corresponding to a free propagator $G_{E,0}$) in any of the diagrams contributing to $\Sigma_E(p)$. Algebraically, this corresponds to considering the function $\delta\Sigma_E(p)/\delta G_{E,0}(q)$ which is shown to be given by:
\begin{equation}\label{eq:coupling_div}
\frac{\delta\Sigma_E(p)}{\delta G_{E,0}(q)}=\frac{V_E(p,q)/2}{(1+G_{E,0}(q)\Sigma_E(q))^2}
\end{equation}
with
\begin{equation}
V_E(p,q)=\Lambda_E(p,q)-\frac12\int_r\!\!\!\!\!\!\Sigma\, V_E(p,r)[G_E(r)]^2\Lambda_E(r,q)
\label{eq:BS}
\end{equation}
and
\begin{equation}
\Lambda_E(p,q)\equiv\lambda+\delta\lambda_E-\lambda^2\int_k\!\!\!\!\!\!\Sigma\, G_E(k)G_E(p-q-k)\,.
\end{equation}
Equation (\ref{eq:coupling_div}) tells that if one wants to properly absorb coupling subdivergences appearing in $\Sigma_E(p)$, one needs to renormalise the function $V_E(p,q)$ as well. This is actually done by means of the following renormalisation condition at the reference temperature $T^\star$:
\begin{equation}\label{eq:ren2}
V^\star_E(p^\star,p^\star)=\lambda\,.
\label{eq:RC2}
\end{equation}
This fixes the value of the coupling counterterm $\delta\lambda_E$ and ensures that $V_E(p,q)$ as well as $\Sigma_E(p)$ are not sensitive to the UV cut-off at any temperature $T$. 

Although it is possible to solve \refeq{eq:BS} exactly as a linear algebra
problem, we recommend the strategy detailed in \refalg{alg:BS}, see App.~\ref{app:alg}. In Ref.~\cite{RenormThermo} we have solved this equation numerically in imaginary time and calculated the renormalised 2PI pressure of a scalar $\varphi^4$ theory to three-loop order in the 2PI-loop expansion. 

\subsubsection{Real-time counterterms}
In principle the counterterms in real time should be equal to those in imaginary time, as long as time is not discretised. We have checked this numerically for $\varphi^4$ theory, but the agreement was
not accurate, unless we used extremely anisotropic regulators ($a_t/a \lesssim
0.05$), which is a highly inconvenient choice for the subsequent
nonequilibrium application due to the expense of the required storage and
computation. If instead, we want to use a conveniently coarse discretisation,
we should not rely on the counterterm values obtained in imaginary time but
rather compute them in real time, directly. Still, calculations in imaginary
time are not totally useless here since they allow for a cheap determination of
real time counterterms (without actually solving a real time version of
Eq.~(\ref{eq:BS})), as we now explain.

First of all, we notice that in imaginary time the field strength counter-terms $\delta Z_t$ and $\delta Z_s$ grow logarithmically with the cutoff with a prefactor of the order of $10^{-5}$. This means that the finite part of the self-energy reaches a cut-off insensitive value before field strength divergences can even show up. In other words, for the cut-off values we use, we can equally drop $\delta Z_s$ and $\delta Z_t$ and still obtain cut-off insensitive results.\footnote{Including the counterterms $\delta Z_t$ and $\delta Z_s$ presents no difficulty in principle.} We assume that this is also true in real time and check this a posteriori. 

We still need to fix the mass and coupling counterterms in real time. The trick we use here to avoid solving the real-time version of Eq.~(\ref{eq:BS}) is that the coupling can equally be fixed by considering any other coupling dependent observable, such as the thermal mass or the renormalisation scale dependence \cite{JakovacRenorm}. Suppose then that we know the thermal mass $m^2_{\rm th}(T)$ from a calculation in imaginary time (we may use here a convenient close-to-continuum regulator, e.g. $a_t\ll a$, $aT\ll
1$, $am\ll1$), defined as
\begin{equation}\label{eq:RC3}
m^2_{\rm th}(T)\equiv
\Sigma_E(p=0)\,,
\end{equation}
where $\Sigma_E(p=0)$ contains the imaginary time counterterms $\delta m_E^2$ and $\delta\lambda_E$, and renormalisation is implemented according to Eqs.~(\ref{eq:ren1}) and
(\ref{eq:ren2}), see Algorithms (\ref{alg:SigmaE}) and (\ref{alg:BS}). We can
evaluate this thermal mass at different temperatures. In particular if we
evaluate it for two different temperatures we have in principle enough
information to recover the values of the counterterms $\delta m_E^2$ and
$\delta\lambda_E$. Now, the thermal mass has also an expression in real-time
which involves the real-time counterterms
\begin{equation}
m^2_{\rm th}(T)=\delta m^2+\frac{\lambda+\delta\lambda}{2}\int\frac{d^4k}{(2\pi)^4}F(k)+\int_0^\infty \!\!\!dt\,\Sigma_\rho(t;\vec{p}=\vec{0})\,.
\end{equation}
 Using the two values for the thermal mass obtained in imaginary-time we can obtain the values for the real-time counterterms. This is used in Algorithm (3), see App.~\ref{app:alg}.

\section{Departing from equilibrium}\label{sec:ooe}

Let us now consider a nonequilibrium situation. If we aim at studying late-time dynamics and thermalisation, the evolution should include the counterterms obtained in the previous section in order to describe the final equilibrated state properly. Because these counterterms do not depend on time, they will be present at any time of the evolution. Then, depending on how one sets up the evolution equations, this might lead
to problems such as a UV divergent time evolution. In Sec.~3.1, we first
quickly revisit the evolution equations with a Gaussian initial condition and
discuss why one fails in obtaining a continuum limit for this type of
evolution. In Sec.~3.2, we start the evolution from a self-consistently
dressed initial quantum state. We give arguments for the existence of a continuum limit in this case and support them with numerical evidence.

\subsection{Gaussian initial condition}
A popular initialisation of the propagator is a Gaussian quantum state
\cite{AartsBergesScalarThermalisation}.
A remarkable feature of this special choice is that the memory integrals
start at initial time, and the only input parameter to be fixed is the
initial propagator and its first derivatives at equal time. The equations of motion in the $F$/$\rho$ formalism read
\cite{AartsBergesScalarThermalisation}:
\begin{eqnarray}
\prn{\partial_x^2\!+\!m^2\!\!+\!\Sigma_{0}(x)}\!F(x,y) &\!\!\!=\!\!\!&
\!\!\int\limits^{y_0}_0 \!\!d^4z\Sigma^F(x,z)\rho(z,y)
\!-\!\!\!\int\limits^{x_0}_0 \!\!d^4z\Sigma^\rho(x,z)F(z,y),\label{eq:Fgauss}\\
\prn{\partial_x^2\!+\!m^2\!\!+\!\Sigma_{0}(x)}\rho(x,y)\! &\!\!\!=\!\!\!&
-\!\!\int\limits^{x_0}_{y_0} \!\!d^4z\Sigma^\rho(x,z)\rho(z,y)\label{eq:rhogauss}\,.
\label{eq:gaussiceom}
\end{eqnarray}

This initial prescription has been frequently used to successfully describe
systems with a fixed cut-off. In low dimensional systems
renormalisation is less problematic \cite{BergesON}.
In a 3+1 dimensional continuum quantum field theory, however,
this class of initial conditions is not useful for the continuum limit of the
time evolution is not accessible. One simple way to view this is to consider
the equation for $F(x,y)$ at initial time $\dbm{x_0=y_0=0}$. The contribution $\Sigma_0(x)$ in the left-hand-side of this equation contains the counterterms which are supposed to renormalise the equilibrium state obtained from the time
evolution. In contrast, the right-hand-side of the equation is equal to zero and
thus, part of the diagrams which should be absorbed by the counterterms are
absent: One has an unbalanced divergence.

\begin{figure}[htbp]
\centerline{\includegraphics[height=3in]{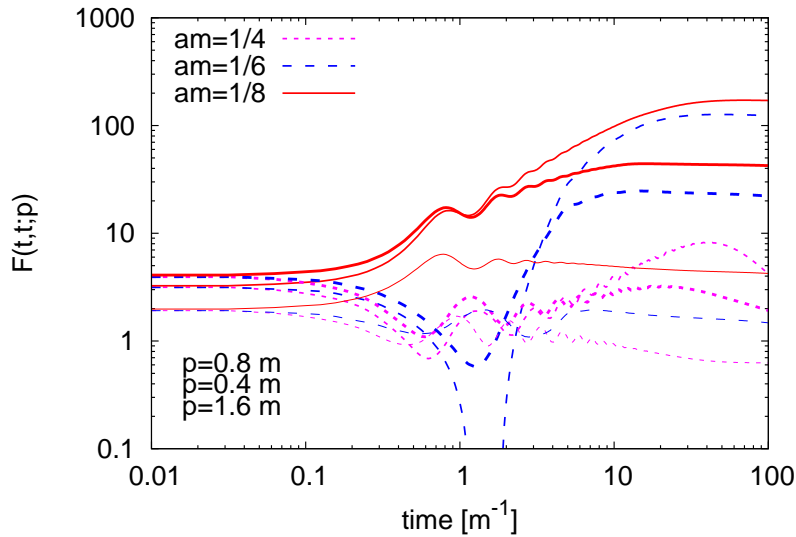}}
\caption{
Time evolution of the equal-time statistical correlator $F(t,t;\vec{p})$ with fixed Gaussian initial propagator ($\dbm{\lambda=24}$, $\dbm{\mbox{box size}: L=32}$) for three different modes $\dbm{|\vec{p}|=0.4,0.8,1.6}$ and three different values of lattice spacing $\dbm{am=1/4,1/6,1/8}$. The counterterms we determine at temperature $\dbm{T^\star=m}$ are not enough to ensure a continuum limit of the time evolution.\label{fig:gauss}
}
\end{figure}

In Fig.~\ref{fig:gauss} we present numerical results for the time evolution
obtained from this kind of initial condition. We parametrise the
initial equal-time statistical correlator $F(t,t;\vec p)$ using the particle number
\begin{equation}
n(\vec p)=N\exp(-(|\vec p|-p_c)^2/2\sigma^2)
\label{eq:nofp}
\end{equation}
with $\dbm{N=5}$, $\dbm{\sigma=0.6m}$, $\dbm{p_c=m}$. We use the counterterms determined at $\dbm{T^\star=m}$ and plot the evolution of three different modes $\dbm{|\vec{p}|=0.4,0.8,1.6}$ for three different values of lattice spacing $\dbm{am=1/4,1/6,1/8}$. On such fine lattices one expects at least an approximate convergence as $am\to0$, as it will indeed happen in the next
subsection. These curves, however, show no sign of a continuum limit. In what
follows, we propose a possible way to cure this.

\subsection{Self-consistent initial conditions}
The general framework of Sec.~\ref{sec:equilibrium} suggests an alternative way to bring the system out of thermal equilibrium. One simply has to calculate the 2PI propagator in presence of a non-vanishing external source $K(x,y)$ living on the real-time path. In this paper we assume the components $K_0(x)$ and $K_\rho(x,y)$ to vanish, in such a way that the equations for the 2PI propagator read
\begin{eqnarray}
\prn{\partial_x^2\!+\!m^2\!+\!\Sigma_{0}(x)}\!F(x,y) & \!\!\!\!=\!\!\! &
\!\!\!\!\int\limits^{y_0}_{-\infty} \!\!dz^4\Sigma_F^K(x,z)\rho(z,y)
\!-\!\!\!\!\int\limits^{x_0}_{-\infty} \!\!dz^4\Sigma_\rho(x,z)F(z,y),\label{eq:FeqK1}\\
\prn{\partial_x^2\!+\!m^2\!+\!\Sigma_{0}(x)}\!\rho(x,y) & \!\!\!\!=\!\!\!\! &
- \!\!\int\limits^{x_0}_{y_0} \!\!dz^4\Sigma_\rho(x,z)\rho(z,y)\,,\label{eq:rhoeqK1}
\end{eqnarray}
where $\dbm{\Sigma_F^K\equiv\Sigma_F+iK_F}$.
The benefit of this approach is that the renormalisability of the evolution
equations only depends on the properties of the source $K(x,y)$. As we motivate
in App.~\ref{app:ext_source}, once the problem has been properly renormalised
in the absence of source, see Sec.~\ref{sec:renormalisation}, introducing a
source does not bring new divergences provided the UV behavior of the source is
well under control.
More precisely we show that, on the imaginary-time path, a source $K$ with proper asymptotics
does not alter the UV structure of the Euclidean theory. To do so we expand the self-energy $\Sigma_E$ in powers of $K$ around the equilibrium solution $\Sigma_E^{K=0}$ which we know how to renormalise:\footnote{Notice that the time evolution corresponding to a Gaussian initial condition,
see previous section, can also be recast in terms of a source and one might wonder why it is not possible to follow the same strategy. The problem in that case is that the propagator at vanishing source has the incorrect (perturbative) asymptotics.
}
\begin{equation}
\Delta\Sigma_E\equiv \Sigma_E-\Sigma_E^{K=0}=\sum_{n\geq 1}\frac{1}{n!}\left.\frac{\delta^n\Sigma_E}{\delta K^n}\right|_{K=0}K^n
\end{equation}
and show that each term of this expansion is finite provided $K$ has the correct UV asymptotics. Of course, strictly speaking, it is not completely obvious that the result of App.~\ref{app:ext_source} obtained on the imaginary-time path can be applied to the real-time path. We shall however provide numerical evidence for a continuum limit on the real-time path which is an indication that our argument does not really depend on the contour we consider.

Suppose we use a source given by
\begin{equation}\label{eq:source}
K_F(x,y)\equiv\left\{\begin{array}{ll}
K_F(x-y)&\textrm{if $x_0<0\,$ and $\,y_0<0$\,;}\\
0&\textrm{if $x_0>0\,\,\,$ or $\,\,\,y_0>0$\,.}
\end{array}\right.
\label{eq:KFswitchoff}
\end{equation}
For times smaller than zero, the system is in a steady state sustained by the
translationally invariant source $K_F(x-y)$. Such states exist for all times
and can be obtained by solving the simplified, translationally invariant
equations
\begin{eqnarray}
\prn{\partial_x^2+m^2+\Sigma_{0}}F(x) & \!\!\!\!=\!\!\!\! &
\!\!\int\limits^{0}_{-\infty} \!\!dz^4\Sigma_F^K(x-z)\rho(z)
-\!\!\int\limits^{x_0}_{-\infty} \!\!dz^4\Sigma_\rho(x-z)F(z)\,,\label{eq:FeqK2}\\
\prn{\partial_x^2+m^2+\Sigma_{0}}\rho(x) & \!\!\!=\!\!\! &
- \int\limits^{x_0}_{0} \!\!dz^4\Sigma_\rho(x-z)\rho(z)\,.\label{eq:rhoeqK2}
\end{eqnarray}
The source $K_F(x-y)$ alone does not completely fix the solution of these equations, as it was also the case
in equilibrium (no source). To define a particular solution we
introduce a non-thermal boundary condition in the form of a generalised
KMS condition
\beq\label{eq:KMSK}
F(p_0;\vec{p})=-i\left(\frac{1}{2}+f(p_0;\vec{p})\right)\rho(p_0;\vec{p})\,,
\eeq
where $f(p_0;\vec{p})$ is an arbitrary function. The ad-hoc use of Eq.~(\ref{eq:KMSK}) has already been suggested in
Ref.~\cite{GreinerScalar} as a recipe to prepare a dressed initial state. We
shall restrict to this type of boundary conditions in what follows. Notice that this generalised KMS condition actually fixes the source to
\begin{equation}
K_F(p)\equiv-\left(\frac12+f(p_0;\vec{p})\right)\Sigma_\rho(p)
+i\,\Sigma_F(p)\,,
\end{equation}
as it can be shown by writing Eqs.~(\ref{eq:FeqK2})-(\ref{eq:rhoeqK2}) in Fourier space.
We need to require that $f(p_0;\vec{p})$ approaches a thermal Bose-Einstein
factor with some temperature sufficiently fast in the UV.
This restriction makes sure that the expansion in $K$ used in the above argument is defined around equilibrium and that the asymptotics of $K$ is appropriate to ensure that the steady solution admits a continuum limit. The details on
how to obtain the steady propagator are given in Algorithm (4), see App.~\ref{app:alg}. 

Now let us turn to the nonequilibrium dynamics. For times greater than zero, one has to solve
\begin{eqnarray}
\prn{\partial_x^2+m^2+\Sigma_{0}}\!F(x,y) & \!\!\!=\!\!\! &
\!\!\int\limits^{y_0}_{-\infty} \!\!dz^4\Sigma_F(x,z)\rho(z,y)
-\!\!\int\limits^{x_0}_{-\infty} \!\!dz^4\Sigma_\rho(x,z)F(z,y)\label{eq:FeqK3}\\
\prn{\partial_x^2+m^2+\Sigma_{0}}\!\rho(x,y) & \!\!\!=\!\!\! &
- \int\limits^{x_0}_{y_0} \!\!dz^4\Sigma_\rho(x,z)\rho(z,y)\,,\label{eq:rhoeqK3}
\end{eqnarray}
where in the integrals, the parts involving times smaller than zero involve the pre-calculated steady solution we just discussed. 
In the course of time evolution $x_0$ is always considered as the most
recent time. Notice, that for practical purposes, we only keep the latest part
of the memory integrals with $|x_0-z_0|<t_{\rm mem}$.
Equations (\ref{eq:FeqK3})-(\ref{eq:rhoeqK3}) look similar to (\ref{eq:Fgauss}) and (\ref{eq:rhogauss}) but
there is in fact a big difference: The memory integrals are present at time
$\dbm{x_0=y_0=0}$ already and can therefore prevent an unbalanced divergence, unlike what happened in the case of a Gaussian initial condition. Still for the divergences to be completely cancelled by the counterterms, the inhomogeneous source $K$, see Eq.~(\ref{eq:source}), should again have the required
asymptotics. This in turn depends on how one chooses $f(p_0;\vec{p})$ in the generalized KMS condition. In this paper we shall not prove analytically that this source has the
correct asymptotic behavior. Rather we shall provide numerical evidence for this, by generating a cut-off
insensitive time evolution.

\begin{figure}[htpb] 
\centerline{\includegraphics[width=4in]{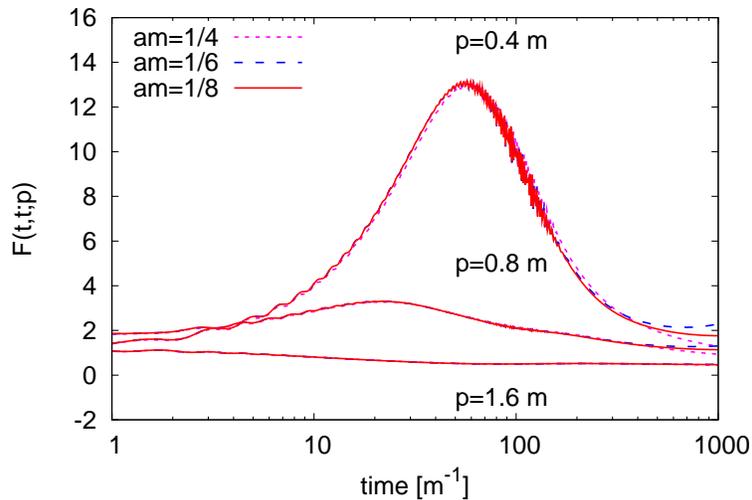}}
\caption{
Time evolution of $F(t,t;\vec p)$ ($\dbm{\lambda=24}$, $\dbm{\mbox{box size}: L=32}$) for three different modes $\dbm{|\vec{p}|=0.4,0.8,1.6}$ and three different values of lattice spacing $\dbm{am=1/4,1/6,1/8}$. The counterterms are determined at temperature $\dbm{T^\star=m}$.
For a given mode, the curves corresponding to runs with a different lattice spacing almost exactly match
each other, which indicates that the continuum limit has been reached. We have checked that the curve for $am=1/2$ still shows sizeable
deviations. 
\label{fig:contlim}
}
\end{figure}

In Fig.~\ref{fig:contlim} we present the time evolution of the equal-time
statistical propagator $F(t,t;\vec p)$ for three different modes $\dbm{|\vec{p}|=0.4,0.8,1.6}$ and three different values of lattice spacing $\dbm{am=1/4,1/6,1/8}$.  We prepare the initial correlated state 
with
\begin{equation}
f(p_0;\vec p)=\frac1{e^{p_0/T(\vec p)}-1}
\end{equation}
where $\dbm{T(\vec p)=T^\star+\omega_{\vec p}/\log(1+1/n(\vec p))}$,
$\dbm{\omega_{\vec{p}}^2\equiv \vec{p}^2+m_{\rm th}^2(T^\star)}$ and
$n(\vec{p})$ is the initial particle distribution (\ref{eq:nofp}) that we
used in the case of the Gaussian initial condition. 
 The counterterms are again determined in equilibrium at temperature $\dbm{T^\star=m}$. For a given mode, the curves representing runs for different values of the spatial cut-off lie almost exactly on top of each other, showing
that the continuum limit has been reached.
This plot strongly suggests that our approach to out-of-equilibrium renormalisation correctly removes all UV divergences and thus allows to define a continuumlimit.

The final relaxation to equilibrium is particularly sensitive to the truncation
of the memory integral. For the plot we used $\dbm{t_{\rm mem}=12m^{-1}}$.
Keeping longer memory we could avoid the small deviation of the curves
at $\sim 1000 m^{-1}$. Long memory integrals, however, are vulnerable to
``instabilities'' on the course of the time evolution. Traces of
such ``instabilities'' can already be seen in Fig.~\ref{fig:contlim} on the
uppermost curves at $\sim 100 m^{-1}$.  This phenomenon manifests in a
high-frequency oscillation on the low-momentum $F(t,t;\vec p)$ curves, and
eventually relaxes. Also, after this high-frequency oscillations have died out,
the $F(t,t;\vec p)$ function continues normally and maintains the approximate
$am$-independence. This phenomenon persisted in an increased volume.
Nevertheless, from our test run with $L=48$ we can conclude that the curves in
Fig. \ref{fig:contlim} are very close to the infinite volume limit up to about
$\sim1000m^{-1}$ where finite volume effects start to matter.

\section{Summary}\label{sec:summary}

We have discussed the renormalisation of the initial value problem in a scalar
quantum field theory. We have introduced a self-consistent initialisation of
the 2PI equations of motion that defines a cut-off independent initial quantum
state. We have also argued for the absence of new divergences in this setting
and performed a numerical analysis to show the continuum limit evolution of a
nonequilibrium quantum field theory.

In this paper we considered the excitation of an originally
equilibrium system using a two-point source. Although we have not discussed
it in this paper, one could give similar arguments for
the finiteness of the solution in presence of a one-point source $J(x)$ with proper asymptotics.
For a spatially homogeneous initial condition, $J(x)$ may only depend on time. This kind of
initialisation is for example suited to describe the parametric resonance scenario.

The primary goal of this paper is to provide recipes. Although scalar fields
have already been extensively discussed out-of-equilibrium in the 2PI approach,
most algorithms and technical details have not been well documented yet.  To
encourage research groups to start working in this field we provide the
detailed algorithms used in this work.

Although scalar theories are interesting on their own (e.g. in cosmological
scenarios), they are often considered as a warm up exercise for the more
complicated gauge theories. The renormalisation of the 2PI effective action for
quantum electrodynamics has been understood by now in equilibrium \cite{RSQEDRenorm}, and a
similar nonequilibrium discussion and numerical analysis is possible. One issue
of interest which could be then studied is the gauge-fixing dependence of the
time evolution.

\subsection*{Acknowledgements}
The authors would like to thank J\"urgen
Berges, Mathias Garny, Markus Michael M\"uller and Julien Serreau for fruitful discussions on related issues. This research was supported in part by the National Science Foundation under
Grant No. PHY05-51164. SB enjoyed the hospitality of the Kavli Institute of
Theoretical Physics. SB was funded by the STFC.

\appendix

\section{Equivalence of contours}\label{app:equivalence}
We would like to show here how to construct any contour-ordered propagator from the real-time propagator defined as the solution $G(x,y)$ to Eqs.~(\ref{eq:SD}) and (\ref{eq:self_2loop}) on the real-time path, obeying the equal time commutation relations and the KMS condition. As we have seen in Sec.~\ref{sec:equiv}, these boundary conditions specify a unique solution. It will be convenient to consider the corresponding Wightman function $G^>(x,y)$, defined, for the moment, for real values of the time arguments. Combining Eqs.~(\ref{eq:fourierFeq}) and (\ref{eq:fourierrhoeq}), one obtains an equation for $G^>(x,y)$ in Fourier space: 
\begin{equation}
(-p_0^2+\omega_{\vec{p}}^2)\,G^>(p)=\int \frac{d\omega}{2\pi}\left[\frac{\Sigma^>(p)\,\rho(\omega;\vec p)}{i(p_0-\omega-i\epsilon)}
+
\frac{\Sigma_\rho(\omega;\vec p)\,G^>(p)}{i(p_0-\omega+i\epsilon)}
\right]\,,
\end{equation}
where $\dbm{\omega_{\vec{p}}^2\equiv \vec{p}^2+m^2+\Sigma_0}$ and $\dbm{\Sigma_0=(\lambda/2)G(0)=(\lambda/2)G^>(0)}$. Using the KMS condition in the form $\dbm{\rho(\omega;\vec{p})=ig(\omega)G^>(\omega;\vec{p})}$, where $\dbm{g(\omega)\equiv 1-e^{-\beta\omega}}$, and the fact that a similar condition holds between the self-energies $\Sigma_\rho(\omega;\vec{p})$ and $\Sigma^>(\omega;\vec{p})$, see Sec.~\ref{sec:equiv}, we arrive at 
\begin{equation}\label{eq:bob2}
\left(-p_0^2+\omega_{\vec{p}}^2\right) G^>(p)=\int\frac{d\omega}{2\pi}\,g(\omega)\left[\frac{\Sigma^>(p)G^>(\omega;\vec{p})}{p_0-\omega-i\varepsilon}+\frac{\Sigma^>(\omega;\vec{p})G^>(p)}{p_0-\omega+i\varepsilon}\right]\,.
\end{equation}\\

We can now use $G^>(p)$ to define $G^>(x,y)$ for complex values of the time arguments:
\begin{equation}\label{eq:analytic}
G^>(x,y)\equiv \int\frac{d^4p}{(2\pi)^4}\,e^{-ip(x-y)}G^>(p)\,.
\end{equation} 
If we assume that $G^>(p)$ does not grow exponentially at large positive frequencies, this definition makes sense provided that $-\beta<{\rm Im}\,(x_0-y_0)<0$. Indeed, if ${\rm Im}\,(x_0-y_0)<0$ the integral is well defined as $\dbm{p_0\rightarrow +\infty}$. Moreover, thanks to the KMS condition which relates the behavior at large positive and negative frequencies, it is easy to see that the condition $-\beta<{\rm Im}\,(x_0-y_0)$ guarantees that the integral is well defined as $\dbm{p_0\rightarrow -\infty}$. Thanks to the analytic continuation (\ref{eq:analytic}), we can now define, on any admissible contour $\mathcal{C}$, the following function
\begin{equation}
G_{\mathcal{C}}(x,y)=\Theta_{\mathcal{C}}(x_0,y_0)G^>(x,y)+\Theta_{\mathcal{C}}(y_0,x_0)G^>(y,x)\,.
\end{equation}
The purpose of such a definition will become clear in a moment. Notice that, due to the equal-time commutation relations, acting twice with $\partial_{x}$ on $G_{\mathcal{C}}(x,y)$ generates a delta function on the contour. Keeping this in mind, one arrives at (we consider here the case $\dbm{{\rm Im}\,x_0<{\rm Im}\,y_0}$)  
\begin{equation}
i(\partial_{x_0}^2+\omega_{\vec{p}}^2)\,G_{\mathcal{C}}(x_0,y_0;\vec{p})=\delta_{\mathcal{C}}(x_0,y_0)+\int\frac{dp_0}{2\pi} e^{-ip_0(x_0-y_0)}i(-p_0^2+\omega_{\vec{p}}^2)G^>(p)\,.
\end{equation}
Using Eq.~(\ref{eq:bob2}) and a simple change of variables, one finally obtains 
\begin{eqnarray}
\delta_{\mathcal{C}}(x_0,y_0) & \!\!\!=\!\!\! & i(\partial_{x_0}^2+\omega_{\vec{p}}^2)\,G_{\mathcal{C}}(x_0,y_0;\vec{p})\nonumber\\
& \!\!\!+\!\!\! & \int\!\frac{d\omega_1}{2\pi}\!\int\!\frac{d\omega_2}{2\pi}\,\Sigma^>(\omega_1,\vec{p})G^>(\omega_2,\vec{p})\nonumber\\
& & \hspace{2.5cm}\times\,\frac{e^{-i\omega_1(x_0-y_0)}g(\omega_2)-e^{-i\omega_2(x_0-y_0)}g(\omega_1)}{i(\omega_1-\omega_2)}\,,
\label{eq:originally55}
\end{eqnarray}
where we have dropped the $\epsilon$ in the denominator since the integrand is well defined as $\omega_1$ and $\omega_2$ become equal.\\

Now we would like to show that $G_{\mathcal{C}}(x_0,y_0;\vec{p})$ is in fact
the contour ordered propagator in the approximation at hand. In other words, we
need to show that $G_{\mathcal{C}}$ fulfills Eqs.~(\ref{eq:SD}) and
(\ref{eq:self_2loop}) on the contour $\mathcal{C}$. To this purpose let us
evaluate the right-hand-side of Eq.~(\ref{eq:SD}). It can be written as
\begin{eqnarray}
R \equiv i(\partial_{x_0}^2+\omega_{\vec{p}}^2)G_{\mathcal{C}}(x_0,y_0;\vec{p}) 
& \!\!\!+\!\!\! & \int_{t_I}^{y_0}dz_0\,\Sigma^>(x_0,z_0;\vec{p}) G^>(y_0,z_0;-\vec{p})\nonumber\\
& \!\!\!+\!\!\! & \int_{y_0}^{x_0}dz_0\,\Sigma^>(x_0,z_0;\vec{p}) G^>(z_0,y_0;\vec{p})\nonumber\\
& \!\!\!+\!\!\! & \int_{x_0}^{t_I-i\beta}\!\!\!\!\!dz_0\,\Sigma^>(z_0,x_0;-\vec{p}) G^>(z_0,y_0;\vec{p})\,,
\end{eqnarray}
where $\dbm{\omega_{\vec{p}}^2=m^2+\Sigma_0}$ with $\dbm{\Sigma_0=(\lambda/2)G_{\mathcal{C}}(0)}$ and $\dbm{\Sigma^>(x,y)=-(\lambda^2/6)G^>(x,y)^3}$. Notice that $\Sigma_0$ coincides with the one in Eq.~(55) because $\dbm{G_{\mathcal{C}}(x,x)=G(x,x)}$. Moreover, because $\Sigma^>(x,y)$ is simply the cube of $G^>(x,y)$, one can use the fact that, for any complex times $x_0$ and $y_0$ such that $\dbm{-\beta<{\rm Im}\,(x_0-y_0)<0}$ 
 \begin{equation}
\Sigma^>(x,y)\equiv \int\frac{d^4p}{(2\pi)^4}\,e^{-ip(x-y)}\Sigma^>(p)\,,
\end{equation} 
to arrive at
\begin{eqnarray}
R & \!\!\!\equiv\!\!\! & i(\partial_{x_0}^2+\omega_{\vec{p}}^2)G_{\mathcal{C}}(x_0,y_0;\vec{p})\nonumber\\ 
& \!\!\!+\!\!\! & \int_{\omega_1,\omega_2}\!\!\!\!\!\!\Sigma^>(\omega_1;\vec{p})\,G^>(-\omega_2;-\vec{p})\,e^{-i\omega_1x_0}e^{i\omega_2y_0}\frac{e^{i(\omega_1-\omega_2)y_0}-e^{i(\omega_1-\omega_2)t_I}}{i(\omega_1-\omega_2)}\nonumber\\
& \!\!\!+\!\!\! & \int_{\omega_1,\omega_2}\!\!\!\!\!\!\Sigma^>(\omega_1;\vec{p})\,G^>(\omega_2;\vec{p})\,e^{-i\omega_1x_0}e^{i\omega_2y_0}\frac{e^{i(\omega_1-\omega_2)x_0}-e^{i(\omega_1-\omega_2)y_0}}{i(\omega_1-\omega_2)}\nonumber\\
& \!\!\!+\!\!\! & \int_{\omega_1,\omega_2}\!\!\!\!\!\!\Sigma^>(-\omega_1;-\vec{p})\,G^>(\omega_2;\vec{p})\,e^{-i\omega_1x_0}e^{i\omega_2y_0}\frac{e^{i(\omega_1-\omega_2)(t_I-i\beta)}-e^{i(\omega_1-\omega_2)x_0}}{i(\omega_1-\omega_2)}\,.\nonumber\\
\end{eqnarray}
Using the KMS condition in the form $\dbm{G^>(-p_0;-\vec{p})=e^{-\beta p_0}G^>(p_0;\vec{p})}$, it is possible to massage the following expression and obtain the right-hand-side of Eq.~(\ref{eq:originally55}). Thus $\dbm{R=\delta_{\mathcal{C}}(x_0,y_0)}$ which proves that $G_{\mathcal{C}}$ is actually the propagator ordered along the contour $\mathcal{C}$, as announced.

\section{Renormalisation with a source}\label{app:ext_source}
Here we would like to study the influence of a source $K$ on the renormalisation of the 2PI self-consistent equations for the propagator. To simplify the discussion we consider here the 2PI effective action on the imaginary-time path. In presence of a source, the Euclidean propagator is given by (space variables implicit)
\begin{equation}
\int_0^\beta d\tau' G_E(\tau,\tau')\left[G_{E,0}^{-1}(\tau',\tau'')+\Sigma_E(\tau',\tau'')+K(\tau',\tau'')\right]=\delta(\tau-\tau'')
\end{equation}
with
\begin{equation}
\Sigma_E(\tau,\tau')=\left.\frac{2\,\delta\Gamma^{E}_{\rm
int}}{\delta G_E(\tau,\tau')}\right|_{G_E}
\end{equation}
where $G_{E,0}$ is the free Euclidean propagator and $\dbm{\Gamma^E_{\rm int}\equiv i\Gamma_{\rm int}}$
is the interaction part of the Euclidean 2PI effective action. We would like to determine under which conditions the counterterms obtained for $\dbm{K=0}$ are enough to renormalise $\Sigma_E$ in presence of the source $K$. In order to compare to the situation in the absence
of source, we consider
\begin{equation}
\Delta\Sigma_E(\tau,\tau')\equiv\Sigma_E(\tau,\tau')-\Sigma^{K=0}_E(\tau,\tau')\,.
\end{equation} 
Assuming that the amplitude of $K$ is small enough, it makes sense to expand
$\Delta\Sigma_E$ as a functional Taylor expansion in powers of $K$. We then
would like to find out under which conditions, the different terms of these
expansion are void of UV divergences. Let us in particular consider the
contribution linear in $K$. At leading order in $K$,
\begin{equation}
\Delta\Sigma_E(\tau,\tau')=-\frac{1}{2}\int_{\eta,\eta',\rho,\rho'}V_E(\tau,\tau;\eta,\eta')G^{K=0}_E(\eta,\rho)G^{K=0}_E(\rho',\eta)K(\rho,\rho')
\end{equation}
with
\begin{eqnarray}
V_E(\tau,\tau';\lambda,\lambda') & \!\!\!=\!\!\! & \Lambda_E(\tau,\tau';\lambda,\lambda')\nonumber\\
& \!\!\!-\!\!\! & \frac{1}{2}\int_{\eta,\eta',\rho,\rho'}\!\!\!\!\!\!\!\!V_E(\tau,\tau';\eta,\eta')G^{K=0}_E(\eta,\rho)G^{K=0}_E(\rho',\eta')\Lambda_E(\rho,\rho';\lambda,\lambda')\,.\nonumber\\
\end{eqnarray}
and
\begin{equation}
\Lambda_E(\tau,\tau';\lambda,\lambda')\equiv\frac{4\,\delta^2\Gamma_{\rm int}^{E}}{\delta G(\tau,\tau')\delta G(\lambda,\lambda')}\,.
\end{equation}

In the previous equations, $G_E^{K=0}$, $\Lambda_E$ and $V_E$ are evaluated at $\dbm{K=0}$, in which case one can use translation invariance. It is then convenient to switch to Fourier space and define
\begin{equation}
G^{K=0}_E(p,p')\equiv(2\pi)^4\delta^{(4)}(p+p')\,G_E(p)
\end{equation}
and
\begin{eqnarray}
\Lambda_E(p,p';q,q') & \!\!\equiv\!\! & (2\pi)^4\delta^{(4)}(p+p'+q+q')\,\Lambda_E(p,p',q)\,,\\
V_E(p,p';q,q') & \!\!\equiv\!\! & (2\pi)^4\delta^{(4)}(p+p'+q+q')\,V_E(p,p',q)\,,
\end{eqnarray}
in terms of which, to linear order in $K$,
\begin{equation}\label{eq:bob}
\Delta\Sigma_E(p,p')=-\frac{1}{2}\int_r\!\!\!\!\!\!\Sigma\;\; V_E(p,p',r)G_E(-r)G_E(p+p'+r)K(-r,p+p'+r)
\end{equation}
with
\begin{eqnarray}\label{eq:BS2}
V_E(p,p',q) & \!\!=\!\! & \Lambda_E(p,p',q)\nonumber\\
& \!\!-\!\! & \frac{1}{2}\int_r\!\!\!\!\!\!\Sigma\;\; V_E(p,p',r)G_E(-r)G_E(p+p'+r)\Lambda_E(-r,p+p'+r,q)\,.\nonumber\\
\end{eqnarray}
This last equation is more general than Eq.~(\ref{eq:BS}). One can, however,
show that the counterterm $\delta\lambda_E$ needed to renormalise
Eq.~(\ref{eq:BS}) is also the one needed to renormalise
Eq.~(\ref{eq:BS2}).\footnote{One can for example consider the difference
$\dbm{\Lambda_E(p,p',q)-\Lambda_E(p,-p,q)}$ and expand in powers of
$\dbm{p'+p}$. Each power of $\dbm{p+p'}$ decreases the degree of
divergence, and the only dangerous $\dbm{(p+p')^0}$ contribution is absent.}
So no new counterterms are needed here and $V_E(p,p',q)$ is automatically UV
convergent. Since $V_E(p,p',q)$ grows at most logarithmically with increasing
momentum $q$ (Weinberg's theorem) and since $G_E(q)$ decreases as $1/q^2$ (up
to logarithms), we conclude from Eq.~(\ref{eq:bob}) that, to linear order in
$K$, $\Delta\Sigma_E$ is UV convergent provided that, for any value of the
fixed momentum $u$, the source $K$ decreases at least like\footnote{Notice that
this condition enforces a particular dependence of $K(p,q)$ as $\dbm{p-q}$
becomes large, for fixed $\dbm{u=p+q}$.}
\begin{equation}\label{eq:asymptotic}
K(p,u-p)\sim\frac{1}{p}
\end{equation}
as $p$ goes to infinity (up to logarithms). A similar analysis shows that, with such a source, higher order contributions to $\Delta\Sigma_E$ (in powers of $K$) are also UV convergent. From this we conclude that provided that the source $K$ follows -- at least -- the asymptotic behavior (\ref{eq:asymptotic}), renormalisation in the absence of $K$ is enough to renormalise the system in the presence of $K$.

\section{Algorithms}\label{app:alg}
The recipes that we disclose here are improved versions of those we used in Ref.~\cite{RenormThermo}. We do not repeat all the numerical
details that we described there.

\begin{alg} \label{alg:SigmaE}
Finding the imaginary-time counterterms $\delta Z_{s,t}$ and $\delta M^2$ by computing the imaginary-time propagator at temperature $T^\star$:
\tt
\begin{enumerate}
\itemsep=0pt
\item Start from $\dbm{G_E^\star=G_0}$.
\item \label{step:SDloop}
      Compute\footnote{
For simple truncations it is advantageous to calculate in direct space: for example, the setting-sun diagram contributes a term $-\lambda^2/6\,G^3_E(x)$ to the self-energy.
} the right-hand-side of Eq.~(\ref{eq:SE}) at $\dbm{T=T^\star}$ keeping the counterterms $\delta Z_{t,s}$ and $\delta M^2$ as parameters.
\item Adjust $\delta Z_{t,s}$ and $\delta M^2$ such that Eqs.~(\ref{eq:RC1}) are satisfied and use them to compute an updated version $G^\star_{E,\rm new}$ of $G^\star_E$.
\item Update $G_E^\star$ using $G_E^\star=\alpha G^\star_{E,\rm new}+(1-\alpha)G^\star_E$ where $\alpha$ denotes\\ a properly adjusted convergence parameter between $0$ and $1$.
\item Iterate from step \ref{step:SDloop} until $G_E^\star$ converges.
\end{enumerate}
\end{alg}
 
\begin{alg}\label{alg:BS}
Finding the imaginary-time counterterm $\delta\lambda_E$ by solving the Bethe-Salpeter equation for $V^\star(p^\star,q)$:\footnote{Since the goal is here to determine $\delta\lambda_E$, it is enough to consider Eq.~(\ref{eq:BS}) for $V_E^\star(p^\star,q)$ at temperature $\dbm{T=T^\star}$. This equation is closed meaning that it does not involve $V_E^\star(p,q)$ at any other value of $p$ but $\dbm{p=p^\star}$.}
\tt
\begin{enumerate}
\itemsep=0pt
\item Once and for all, calculate $\Lambda_E(p,q)$ at $\dbm{T=T^\star}$.
\item Start from $\dbm{V_E^\star(p^\star,q)=\lambda}$.
\item \label{step:BSloop}
      Compute the right hand side of \refeq{eq:BS} for $\dbm{T=T^\star}$ and $\dbm{p=p^\star}$, keeping the counterterm $\delta\lambda_E$ dependence as a parameter. 
      \item Solve the linear algebraic equation (\ref{eq:ren2}) for $\delta\lambda_E$ and use the latter to define an uptaded $V_E^\star(p^\star,q)$.
\item Iterate from step \ref{step:BSloop} until $V_E^\star(p^\star,q)$ converges.
\end{enumerate}

{\rm After fixing $\delta\lambda_E$ it is not necessary to recalculate
$G_E^\star$ with the new counterterms. Neither $\delta Z_{t,s}$ nor $\delta M^2$ are changed. The value of $\delta\lambda_E$ is important to obtain UV convergent results at temperatures away from $T^\star$.}

\end{alg}

\begin{alg}\label{alg:Sigmarho}
Finding the real time counterterms and propagator simultaneously: 
\tt
\begin{enumerate}
\itemsep=0pt
\item Start with $\dbm{\Sigma^\rho(t;\vec p)=0}$.
\item \label{step:rholoop} Loop over the three-momentum $\vec p$:
\begin{enumerate}
\item Initialise $\dbm{\rho(t=0;{\vec p})=0}$, $\dbm{\rho(t=a_t;{\vec p})=a_t}$.
\item \label{step:rhoeq}
Solve the time-explicit ODE (\ref{eq:rhoeq}) for $\rho(t;{\vec p})$
with the thermal mass squared on the left hand side.\footnote{Solve
it over a long time interval (few times relaxation time),
but you can truncate the memory integral at $\sim 10 m^{-1}$, where $m$ is the
relevant mass scale, since
$\Sigma^{\rho}(t;\vec{p})$ relaxes on the microscopical scale. $\Sigma^{\rho}(t;\vec{p})$
is strongly sensitive to volume for $t>L/2$, $L$ being the physical box size.
This also sets a limit to the memory it is meaningful to keep.}
\item \label{step:kms} Deduce $F(t;{\vec p})$ using Fourier transforms and Eq.~(\ref{eq:KMS}).\footnote{The Fourier transform  requires
the knowledge of $\rho(t;\vec{p})$ over an infinite period. In practice, $F(t,\vec p)$ will be accurate enough in the initial period if we have $\rho(t;\vec{p})$ over a few relaxation times only. }
\item Store $\rho(t;\vec p)$ and $F(t;\vec p)$ for the relevant $t$-range.
\end{enumerate}
\item Update the self-energy $\Sigma^{\rho}(t;\vec p)$ and adjust $\delta M^2$ to
\begin{equation}
\delta M^2=-\int_0^\infty\Sigma^{\rho}(t;\vec{p}=\vec{0})\,.\nonumber
\end{equation}
\item
Iterate from step \ref{step:rholoop}.
\item
Repeat this procedure at an other temperature and with the corresponding
thermal mass and use \refeq{eq:RC3} to obtain $\delta\lambda$ and $\delta m^2$.
\end{enumerate}
\end{alg}

\begin{alg}\label{alg:KF}
Generating a far-from-equilibrium self-consistently correlated quantum state (modes are populated according to a given mode temperature $T({\vec p})$): 
\tt
\begin{enumerate}
\itemsep=0pt
\item Start with a $\Sigma^\rho(t;\vec p)$ as obtained in Algorithm (3).
\item \label{step:rholoopKF} Loop over the three-momentum $\vec p$:
\begin{enumerate}
\item Initialise $\rho(t=0;{\vec p})$, $\rho(t=a_t;{\vec p})=a_t$.
\item \label{step:rhoeqKF}
Solve the time-explicit ODE (\ref{eq:rhoeq}) for $\rho(t;{\vec p})$.
\item Deduce $F(t;{\vec p})$ using Fourier transforms and
\begin{equation}
F(\omega;\vec{p}) = -i \left(1/2+1/\left[e^{\omega/T(\vec p)}-1\right]\right)\rho(\omega;\vec{p})\,.\nonumber
\end{equation}
\item Store $\rho(t,\vec p)$ and $F(t,\vec p)$ for the relevant $t$-range.
\end{enumerate}
\item Calculate the self-energy
and iterate from step \ref{step:rholoopKF}.
\end{enumerate}
\end{alg}

\begin{alg}\label{alg:evol}
Solving the nonequilibrium evolution starting from a self-consistent
initial condition.
\tt
\begin{enumerate}
\itemsep=0pt
\item
Generate a translation invariant $F_s(t;\vec p)$ and
$\rho_s(t;\vec p)$ using \refalg{alg:KF}.
\item
Set $\dbm{F(t_x,t_y;\vec p)=F_s(t_x-t_y;\vec p)}$ and
$\dbm{\rho(t_x,t_y;\vec p)=\rho_s(t_x-t_y;\vec p)}$ for
   $-t_{\rm mem}<t_x,t_y<0$.
Set the evolution time to $t=0$.
\item \label{step:evolloop}
Calculate the self energies $\Sigma_{F,\rho}(t,t_y,\vec p)$ for
$t-t_{\rm mem}<t_y\le t$.
\item Use the discretised Eqs.~(\ref{eq:FeqK3})-(\ref{eq:rhoeqK3}) to obtain
$F(t+\delta t,t_y)$ and $\rho(t+\delta t,t_y)$ for $t_{\rm mem}<t_y\le t$.
\item Use the newly calculated propagator values in
Eq.~(\ref{eq:FeqK3}) to obtain $F(t+\delta t,t+\delta t)$.
\item Increase $t$ by $\delta t$ and repeat from step
\ref{step:evolloop}.
\end{enumerate}
\end{alg}

\section{Discretisation of the isotropic propagator}\label{app:isotropy}

It is very natural to start from a discretised action before
defining the 2PI effective action. The straightforward lattice discretisation
of a homogeneous system will then lead to a propagator $G(t_1,t_2,\vec x)$
where $\vec x$ is a lattice 3-vector. The isotropy in $\vec x$ reduces to 
a permutation symmetry between the components of $\vec x$. Using also the
reflection symmetry one can conclude that if the action is defined on
a lattice with an even linear size $N$, the total number of variables that
describes a propagator at a given pair of time coordinates is
$(N+6)(N+4)(N+2)/48$. This discretisation has been used by several groups,
starting with Ref.~\cite{HeidelbergParametric},
and also in Ref.~\cite{RenormThermo}.

We now introduce a different kind of discretisation, where the rotation
symmetry of the propagator is exact, and the number of independent variables
in the propagator is $N$. This corresponds to a discretisation
on the level of the 2PI effective action, too.

In many cases we have to calculate a loop integral, which is the
convolution of two symmetric functions
\begin{equation}
(f*g)(p)=\int \frac{d^3k}{(2\pi)^3} f(|\vec k|)g(|\vec p-\vec k|)=
\frac1p\int_0^\infty\!\frac{dk}{2\pi}\, k f(k)
\int_{|p-k|}^{p+k}\!\frac{dq}{2\pi}\, q g(q)\,.
\label{eq:convolution}
\end{equation}
(Here and in the following we use plain letters for the modulus of a lattice
3-vector.)
This convolution is not defined in the strict sense without specifying
a UV regulator. By fixing a momentum-space cut-off one also introduces
Umklapp processes: if an outgoing momentum after a scattering process
lies off the Brillouin zone, one maps this back to one of the modes
inside. We motivate our cut-off prescription with numerical convenience.
We would like to calculate the convolution in Eq.~(\ref{eq:convolution})
as a product co-ordinate space:
\begin{equation}
\tilde{(f*g)}(x)=\tilde f(x) \tilde g(x)\,,
\label{eq:xproduct}
\end{equation}
where the $\tilde f(x)$ is the inverse Fourier transform of $f(k)$ 
in terms of
\begin{eqnarray}
x \tilde f(x) &\!\!\!=\!\!\!& \int_0^\infty \frac{dk}{2\pi^2} k f(k) \sin kx\,.
\label{eq:contisinft}
\end{eqnarray}
We will have to discretise this inverse sine transform
in momentum and co-ordinate.
To avoid problems with the potential singularities at $x\to 0$ and $k\to0$
we choose a discretisation without these points. For the infrared behaviour
will assume that $kf(k)\to0$ as $k\to 0$, which is certainly true in a
massive theory. For algorithmic convenience we extend the momenum space
functions beyond the cut-off $\Lambda$ by the equation
$kf(k)=(2\Lambda-k)f(2\Lambda-k)$, this implicitely determines the Umklapp
behaviour.
 
The discretised sine and inverse sine transforms matching
these boundary conditions are known as DST-II and DST-III formulae,
respectively, these are available in many numerical libraries \cite{FFTW}.
We introduce a lattice spacing $a$, which is related to the
the highest stored momentum by $\Lambda=\pi/a$, the space points
where we will have to use Eq.~(\ref{eq:xproduct}) are
$x_n=a\left(n+\frac12\right)$, with $n=0..N-1$. In momentum space
we define the momentum grid as $k_j=(j+1)\Lambda/N$, with $j=0..N-1$.
Since on a one-dimensional lattice of physical size $L$ the momentum
space lattice spacing is $2\pi/L$, the corresponding "box size" in our
discretisation is $L=2aN$. With these notations the transformation rules
read:

\begin{eqnarray}
f_k&\!\!\!=\!\!\!&4a^3\frac{N}{k+1} \left[
\sum_{n=0}^{N-1} \left(n+\frac12\right)\tilde f_n
\sin\left(\frac{\pi}{N} \left(n+\frac12\right)(k+1)\right)
\right]\,,\\
\tilde f_n&\!\!\!=\!\!\!&\frac{1}{2N^2a^3(n+\frac12)} \left[
\frac{(-1)^n}2Nf_{N-1}\right.\nonumber\\
&&\qquad\qquad\qquad+\left.\sum_{k=0}^{N-2} \left(k+1\right)f_k
\sin\left(\frac{\pi}{N} \left(n+\frac12\right)(k+1)\right)
\right]\,.
\end{eqnarray}

As we have already mentioned, the zero momentum and the equal time
propagators are not stored at all, so
we need separate formulae for the volume integrals:

\begin{eqnarray}
\int_0^\Lambda\frac{d^3k}{(2\pi)^3} f(k)&\!\!\!=\!\!\!&
\frac{\pi}{2(aN)^3}\left[ \frac{N^2}{2}f_{N-1}
+\sum_{k=0}^{N-2}(k+1)^2f_k\right]\,,\\
\int_0^{aN}d^3x \tilde f(x)&\!\!\!=\!\!\!&
4\pi a^3\sum_{n=0}^{N-1}\left(n+\frac12\right)^2\tilde f_n\,.
\end{eqnarray}

We used this kind of cut-off discretisation in our presented numerics.
In the production runs the time-like grid was typically four times finer
than the spatial lattice.


\end{document}